\documentclass[aps,prl,amsmath,twocolumn,amssymb]{revtex4}%
\usepackage{graphicx}
\usepackage{pdfpages}
\usepackage{dcolumn}
\usepackage{bm}
\usepackage{amsmath}
\usepackage{fancybox}
\usepackage{amsfonts}
\usepackage{hyperref}
\usepackage{epstopdf}
\usepackage{amssymb}%

\begin{document}
\title{Controlling Chiral Domain Walls in Antiferromagnets Using Spin-Wave Helicity}
%%authors
\author{Alireza Qaiumzadeh, Lars A. Kristiansen, and Arne Brataas}
\affiliation{Center for Quantum Spintronics, Department of Physics, Norwegian University of Science and Technology, NO-7491 Trondheim, Norway}

\begin{abstract}
In antiferromagnets, the Dzyaloshinskii-Moriya interaction lifts the degeneracy of left- and right-circularly polarized spin waves. This relativistic coupling increases the efficiency of spin-wave-induced domain-wall motion and leads to higher drift velocities. We demonstrate that, in biaxial antiferromagnets, the spin-wave helicity controls both the direction and the magnitude of the magnonic force on chiral domain walls. In this case it is shown that the domain-wall velocity is an order of magnitude faster in the presence of the Dzyaloshinskii-Moriya interaction. In uniaxial antiferromagnets, by contrast, the magnonic force is always propulsive but its strength is still helicity dependent.
\end{abstract}
%\pacs{
%75.50.Ee, % Antiferromagnetics
%75.70.Tj, % Spin-orbit effects
%75.60.Ch, % Domain walls and domain structure
%75.30.Ds  % Spin waves
%      }

\date{\today}
\maketitle

The Dzyaloshinskii-Moriya interaction (DMI) was introduced to describe the weak ferromagnetism in antiferromagnetic (AFM) materials \cite{DMI1,DMI2}. The DMI arises from the relativistic spin-orbit coupling that occurs when there is a broken inversion symmetry. The DMI is an antisymmetric exchange interaction between two adjacent spins $\bm{S}_{i-1}$ and $\bm{S}_{i}$, $w_{\mathrm{DM}}=-\sum_{i} (-1)^i \bm{D}_0\cdot(\bm{S}_{i-1}\times \bm{S}_{i})$, where $\bm{D}_0$ is a vector with constant direction and amplitude \cite{papa}. The direction of $\bm{D}_0$ is dictated by the point group symmetry of materials. The DMI can originate in the bulk in non-centrosymmetric crystals. There also can be interfacial DMIs in ultrathin films and at interfaces with heavy metals \cite{Fert-iDMI, Bogdanov-DMI1,Bogdanov-DMI2, Bogdanov-DMI3}. The discovery of magnetic skyrmions and chiral domain walls (CDWs) \cite{Sky1,Sky2,Sky3,Sky4,CDW, Analisa}  has led to renewed interest in the DMI. In addition to their fundamental interest, the control of these intriguing textures may have applications in next-generation information storage and processing devices.

% A bulk-like DMI has a DM vector that is parallel to the displacement between the adjacent spins. At an interface, the DM vector is perpendicular to the distance between neighboring spins.

AFM materials are ordered magnetic materials without any net magnetization. Their lack of stray fields and THz response are promising for novel ultra-dense and ultra-fast magnetic devices. Topics in AFM spintronics include spin angular momentum transport and transfer, spin-orbit coupling, and the manipulation of AFM domains and solitons  \cite{AFM-light, AFM-Jungwirth, AFM-Manchon, AFM-vortex1, AFM-Sky1,Ivanov-AFM1, Ivanov-AFM2,Ivanov-AFM3,Alireza1, Alireza2}. Coherent or thermal spin waves (SWs) \cite{AFMDW-SW1,AFMDW-SW11, AFMDW-SW2, AFMDW-thermal} and currents
\cite{Alireza1, AFMDW-SOT1, AFMDW-SOT2} can induce movement of AFM domain walls (AFM-DWs). In the absence of the DMI, circularly polarized SWs push AFM-DWs via the transfer of linear momentum \cite{AFMDW-SW1,AFMDW-SW11, AFMDW-SW2} but pull ferromagnetic (FM) domain walls (FM-DWs) via the transfer of spin angular momentum \cite{FMSW1,FMSW2,FMDW-DMI}. The SWs in FM systems are always right-handed. By contrast, in AFM systems, both left- (L) and right-circularly (R) polarized SWs exist \cite{AFMSW1,AFMSW2,AFMSW3}.
In the absence of the DMI, the dispersions associated with the two helicities are degenerate, but the DMI lifts this degeneracy \cite{AFMSW4-DMI}.

The helicity degree of freedom in AFM-SWs, among other degrees of freedom in the nature, such as the spin of electrons and polarization of light, recently has attracted great attention \cite{AFMSW2, AFMSW3, AFMSW4-DMI,helicity1,helicity2,helicity3, helicity4} and promises many applications in novel low dissipation magnonic data processing.

In this Rapid Communication, we demonstrate the possibility of all-magnonic helicity-dependent AFM-DW motion in the presence of DMI. This paves the way for faster and improved control of AFM textures.

\textit{Model.---} We consider a two-sublattice AFM insulator with equal spins, $S_A=S_B=S$. The unit vectors along the directions of the magnetic moments are $\bm{m}_A(\bm{r},t)=\bm{S}_A/S$ and $\bm{m}_B(\bm{r},t)=\bm{S}_B/S$. At equilibrium, $\bm{m}_A(\bm{r},t)$ and $\bm{m}_B(\bm{r},t)$ are antiparallel. We introduce the magnetization $\bm{m}=(\bm{m}_A+\bm{m}_B)/2$ and the staggered order parameter $\bm{n}=(\bm{m}_A-\bm{m}_B)/2$, where $\bm{n}\cdot\bm{m}=0$ and $\bm{n}^2+\bm{m}^2=1$. We consider an effective one-dimensional (1D) model along the $x$ direction to describe the AFM-DW motion.

In the exchange approximation, the total Lagrangian density of an AFM system is
\begin{align}
\mathcal{L}[\bm{n},\bm{m}]=\mathcal{L}_{\mathrm{B}}[\bm{n},\bm{m}]-U[\bm{n},\bm{m}],
\label{lagrangian}
\end{align}
where the Berry-phase-induced term $\mathcal{L}_{\mathrm{B}}$ and the free energy density $U$ are \cite{Ivanov-AFM1, Erlend,Lifshitz,energy-terms}
\begin{subequations}
\begin{align}
&\mathcal{L}_{\mathrm{B}}[\bm{n},\bm{m}]=\rho\partial_t{\bm{n}}\cdot(\bm{n}\times\bm{m}),\label{Lagrangian-nm}\\
&U[\bm{n},\bm{m}]=\frac{\bm{m}^2}{2\chi}+L \bm{m}\cdot \partial_x \bm{n}+\frac{A}{2}(\partial_x \bm{n})^2-\frac{K_x}{2}(\bm{n} \cdot \hat{x})^2\nonumber\\&+w_{DM}.
\label{free-energy-nm}
\end{align}
\end{subequations}
Here, $\rho=\hbar S/a $ is the spin angular momentum density, $a$ is the lattice constant, $\chi$ is the magnetic susceptibility, $A$ is the exchange stiffness, $L$ is the parity-breaking term amplitude \cite{Erlend}, and $K_x > 0$ is the easy axis anisotropy energy density. The DM free energy density in the continuum model,  $w_{\mathrm{DM}}=\bm{d}\cdot(\bm{m}\times \bm{n})+\bm{D}\cdot(\partial_x\bm{n}\times \bm{n})$, consists of homogeneous and inhomogeneous DMIs as expressed by the first and second terms, respectively.
The magnetization is a slave variable and can be found by solving the equation of motion, $\chi^{-1}\bm{m}=\rho \partial_t{\bm{n}} \times \bm{n}-L \partial_x \bm{n}+\bm{d}\times\bm{n}$.

To simplify the expressions, we use natural units of time, length and energy: $t_0=\rho\sqrt{\chi/K_x}$, $\lambda_0=\sqrt{A/K_x}$, and $\epsilon_0=\sqrt{A K_x}$, where the exchange stiffness and easy axis anisotropy are renormalized as follows: $A \rightarrow A-\chi L^2$ and $K_x \rightarrow K_x-\chi d^2$. In this way, the contributions to the Lagrangian density as functions of the staggered order parameter read simply as
\cite{topological-term}
\begin{subequations}
\begin{align}
&\mathcal{L}_{\mathrm{B}}[\bm{n}]=\frac{1}{2}(\partial_t\bm{n})^2, \label{Lagrangian-n}\\
&U[\bm{n}]=\frac{1}{2}(\partial_x\bm{n})^2-\frac{1}{2}(\bm{n} \cdot \hat{x})^2+D\bm{n}\cdot(\hat{x}\times \partial_x \bm{n}).
\label{free-energy-n}
\end{align}
\end{subequations}
The kinetic part of the AFM Lagrangian, Eq.\ \eqref{Lagrangian-n}, is Lorentz invariant \cite{Ivanov95}, and the effective velocity of light is the maximum velocity of magnons in isotropic systems, $c=1$. As in ferromagnets \cite{Oleg-DMI}, Eq.\ (\ref{free-energy-n}) implies that when $D > 1$, the ground state is a helical state with a spatial period of $2\pi/D$. By contrast, when $D<1$, there are two collinear, degenerate ground states: $\bm{n}_0=\sigma \hat{x}$, where $\sigma=\pm$. A domain wall (DW) is a transition between these two discrete degenerate ground states \cite{Kardar}. In the following, we assume that $D$ is smaller than the critical value of $D<1$.

By minimizing the free energy Eq. (\ref{free-energy-n}) with respect to the boundary conditions $\bm{n}_0(x\rightarrow\pm \infty)=\sigma \hat{x}$, we can find the profile of a CDW. This profile is represented by $\bm{n}_0=(\cos\theta,\sin\theta\cos\phi,\sin\theta\sin\phi)$ with $\cos\theta=\tanh[(x-x_0)\sqrt{1-D^2}]$ and $\phi=(x-x_0)D+\Phi$, where $\sqrt{1-D^2}$ is the effective DW width, $D$ is the rate of DW twisting, $\Phi$ is the DW tilt, and $x_0$ is the position of the DW center \cite{Oleg-DMI}. When $D\rightarrow 0$, the DW profile is of the N\'{e}el type. In the limit of $D\rightarrow 1$, the system gradually approaches a spiral state. The energy of CDW is $E=2\sqrt{1-D^2}$.

%$(\hat{\mathrm{e}}_1=\sigma \hat{y},\hat{\mathrm{e}}_2=\hat{z},\hat{\mathrm{e}}_3=\sigma \hat{x})$

\textit{DW motion in a uniaxial AFM system.---} We compute the SW dispersions in uniaxial uniform domains and CDWs $\omega_{\mathrm{u}}$ and $\omega_{\mathrm{c}}$, respectively. In doing so, we assume that there is a small transverse deviation of the staggered field on top of a static ground state $\bm{n}=\bm{n}_0(x)+\delta\bm{n}$. It is convenient to use a global basis to express the SWs \cite{AFMDW-SW11,AFMDW-SW2}. The orthogonal unit vectors are $\hat{\mathrm{e}}_1=\partial \bm{n}_0/\partial\theta$, $\hat{\mathrm{e}}_2=(\partial \bm{n}_0/\partial\phi)/\sin\theta$, and $\hat{\mathrm{e}}_3=\hat{\mathrm{e}}_1 \times \hat{\mathrm{e}}_2=\bm{n}_0$. A SW is represented by a complex field  $\psi=\delta\bm{n}\cdot(\hat{\mathrm{e}}_1+i\hat{\mathrm{e}}_2)$ \cite{SW-expression}. We obtain the effective Lagrangian density for SWs by expanding the total Lagrangian to the second order in $\psi$. The effective SW equation is $\mathcal{H}_{\mathrm{u}(\mathrm{c})}\psi={\omega^2_{\mathrm{u}(\mathrm{c})}}\psi$. The SW Hamiltonians in a uniform domain and in a CDW are
\begin{subequations} \label{SW-Hamil}
\begin{align}
&\mathcal{H}_{\mathrm{u}}=-\partial_x^2+i2\sigma D \partial_x+1, \label{SWH-col}\\
&\mathcal{H}_{\mathrm{c}}=-\partial_x^2+(1-D^2)\left[1-2\mathrm{sech}^2(x\sqrt{1-D^2})\right]. \label{SWH-DW}
\end{align}
\end{subequations}
The eigenvalues are given by
\begin{subequations} \label{SW-disp}
\begin{align}
&\omega^2_{\mathrm{u}}=1-2\sigma D k+k^2, \label{SW-col}\\
&\omega^2_{\mathrm{c}}=1-D^2+k^2, \label{SW-DW}
\end{align}
\end{subequations}
and the eigenfunctions are expressed as $\psi(t,x)=\Psi \psi_0(x) e^{-i\omega_{\mathrm{u}(\mathrm{c})} t + i k x}$, where $\Psi=|\delta\bm{n}|$ is the SW amplitude far from the AFM-DW and $k$ is the wave number.
The eigenfunction $\psi$, together with its dispersion relation as given in Eq. (\ref{SW-disp}), describes a circularly polarized SW. The dispersion relation of Eq.\ \eqref{SW-disp} has both positive- and negative-frequency solutions. The SWs with $\omega<0$ ($\omega>0$) are right-circularly (left-circularly) polarized SWs. Since the phase velocity is $\omega/k$, the left-circularly polarized SWs are rightward-moving for $k>0$ and leftward-moving for $k<0$, whereas the opposite is true for the right-circularly polarized SWs.

In AFM materials, the DMI lifts the degeneracy of the circularly polarized SWs, as has recently been observed experimentally \cite{AFMSW4-DMI}. Equation (\ref{SW-col}) shows that the SW spectrum is non-reciprocal in uniaxial AFM domains, such that $\omega_{\mathrm{u}}(k)\neq \omega_{\mathrm{u}}(-k)$. However, in chiral AFM-DWs, the dispersion is symmetric and degenerate, such that $\omega_{\mathrm{c}}(k)= \omega_{\mathrm{c}}(-k)$, with a renormalized energy gap.
The group velocity in a uniform domain is $v_\mathrm{u}=\left(k-\sigma D\right)/\omega_{\mathrm{u}}$, whereas in a CDW, it is $v_\mathrm{c}=k/\omega_{\mathrm{c}}$.

\begin{figure}[t]
\includegraphics[width=8.5cm]{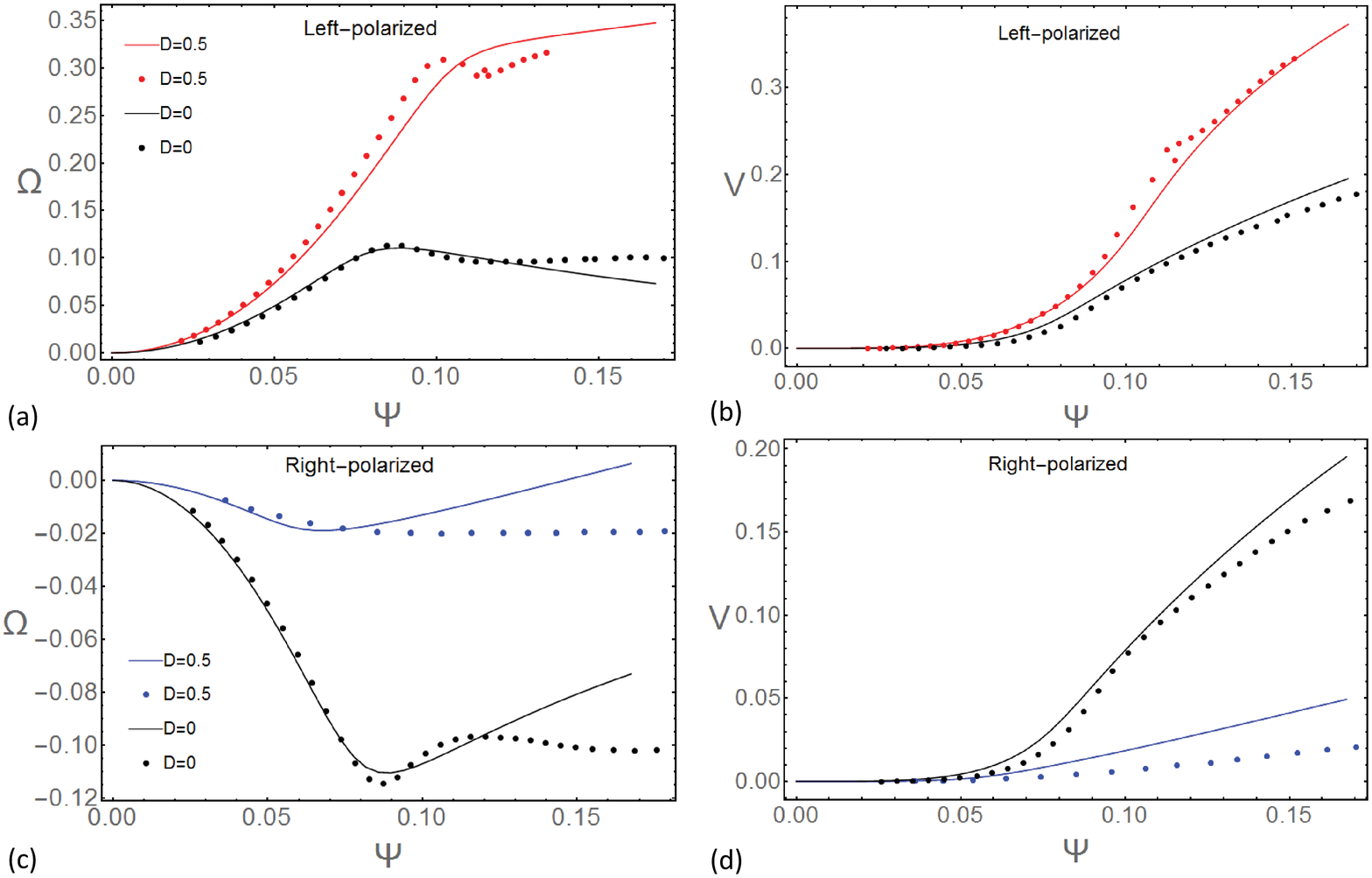}
\caption{The angular frequency $\Omega$ [(a) and (c)] and linear velocity $V$ [(b) and (d)] of an AFM-DW as functions of the SW amplitude $\Psi$ for $k = 0.8$. The solid lines are based on Eqs. (\ref{ref-tran}) and (\ref{reactiveforce})-(\ref{visc}), whereas the circles represent solutions of LLG equations.
The black curves are the results in the absence of DMI and the red/blue curves show the results for a finite DMI in the presence of left-/right-circularly polarized SWs.}
\label{k08}
\end{figure}

The effective potential energy of a \textit{static} AFM-DW, the P\"{o}schl-Teller potential in the SW Hamiltonian $\mathcal{H}_{\mathrm{c}}$ of Eq. (\ref{SWH-DW}), is reflectionless. Nevertheless, a circularly polarized SW exerts a torque on an AFM-DW. As a result, the DWs in a uniaxial AFM material rotate and no longer remain reflectionless \cite{AFMDW-SW1,AFMDW-SW11, AFMDW-SW2}. Consequently, there is a force on the AFM-DWs; see Fig. \ref{k08}.

To capture these effects, we transform from the laboratory frame $\mathcal{F}$ into a uniformly moving and precessing frame $\mathcal{\tilde{F}}$. The new frame of reference $\mathcal{\tilde{F}}$ rotates with an angular frequency of $\Omega$ around the $x$ axis and moves with a linear velocity of $V$ along the $x$ direction \cite{AFMDW-SW11,AFMDW-SW2}. In the new frame, the profile of a CDW is described by $\cos\theta=\tanh[(\tilde{x}-\tilde{x}_0)\sqrt{1-\tilde{D}^2}]$ and $\phi=(\tilde{x}-\tilde{x}_0)\tilde{D}+\Phi$, where we mark the variables in $\mathcal{\tilde{F}}$ with a tilde. The natural units in $\mathcal{\tilde{F}}$ are $\lambda_r=\Gamma\lambda_0$, $t_r=\Gamma t_0$, and $\epsilon_r=\epsilon_0/\Gamma$, where $\Gamma=1/\sqrt{1-\Omega^2+2 \gamma V D \Omega}$ and $\gamma=1/\sqrt{1-V^2}$. Additionally,  we define $\tilde{\Omega}=\Gamma(\Omega-\gamma VD)$ and $\tilde{D}=\Gamma\gamma D$. Finally, in $\mathcal{\tilde{F}}$, the Lagrangian density is given by
\begin{subequations}
\begin{align}
&\tilde{\mathcal{L}}_B=\frac{1}{2}(\partial_{\tilde{t}}\bm{n})^2+\tilde{\Omega} \partial_{\tilde{t}}\bm{n}\cdot(\hat{x}\times\bm{n}), \label{Lagragian-B-newframe} \\
&\tilde{U}=\frac{1}{2}(\partial_{\tilde{x}}\bm{n})^2-\frac{1}{2}(\bm{n}\cdot \hat{x})^2+\tilde{D} \bm{n}\cdot(\hat{x}\times\partial_{\tilde{x}}\bm{n}).\label{Lagrangian-rot}
\end{align}
\end{subequations}
The second term in the kinetic part of the Lagrangian, Eq. \eqref{Lagragian-B-newframe}, is due to the Coriolis force \cite{AFMDW-SW2}. By expanding the effective Lagrangian in the new reference frame $\tilde{\mathcal{L}}$ up to the second order in $\psi$ and calculating the Euler-Lagrange equation, we find an eigenproblem expressed as $\tilde{\mathcal{H}}_{\mathrm{c}}\psi=\tilde{\omega}^2 \psi$, with the following effective Hamiltonian for SWs on a precessing CDW:
\begin{align}
\tilde{\mathcal{H}}_{\mathrm{c}}=& \mathcal{H}_{\mathrm{c}}[x \rightarrow \tilde{x}, D \rightarrow \tilde{D} ]
+2\tilde{\omega}\tilde{\Omega} \tanh(\tilde{x}\sqrt{1-\tilde{D}^2}).\label{Hamiltonian-rot}
%-\partial^2_{\tilde{x}}+(1-\tilde{D}^2)\left(1-2 \mathrm{sech}^2(\tilde{x}\sqrt{1-\tilde{D}^2})\right)\nonumber\\&
%+2\tilde{\omega}\tilde{\Omega} \tanh(\tilde{x}\sqrt{1-\tilde{D}^2}). \label{Hamiltonian-rot}
\end{align}
The additional hyperbolic tangent potential energy in the Hamiltonian $\tilde{\mathcal{H}}_{\mathrm{c}}$ causes reflections \cite{AFMDW-SW11,AFMDW-SW2}.

The reflection $|r|$ and transmission $|t|$ amplitudes for the 1D hyperbolic tangent scattering potential are known \cite{Landau}:
\begin{subequations}\label{ref-tran}
\begin{align}
&|r|^2=\frac{\sinh^2[\frac{\pi}{2}(\tilde{k}_{+}-\tilde{k}_{-})]}{\sinh^2[\frac{\pi}{2}(\tilde{k}_{+}+\tilde{k}_{-})]},\\
&|r|^2\tilde{k}_{-}+|t|^2\tilde{k}_{+}=\tilde{k}_{-}.
\end{align}
\end{subequations}
In $\mathcal{\tilde{F}}$, the wave numbers of the transmitted and incoming SWs $\tilde{k}_{+}$ and $\tilde{k}_{-}$, respectively, are given by $\tilde{k}^2_{\pm}=\tilde{\omega}^2\mp\tilde{\omega}\tilde{\Omega}-1+\tilde{D}^2$. The wave numbers are the same in $\mathcal{F}$ and $\tilde{\mathcal{F}}$; thus, ${k}^2_{\pm}=\omega_\pm^2-1+D^2$. The frequency shift between an incoming SW of frequency $\omega_-=\omega+(\Omega-\gamma V D)$ and an outgoing SW of frequency $\omega_+=\omega-(\Omega-\gamma V D)$ is our first central result:
\begin{align}
\Delta\omega=-2(\Omega-\gamma V D).
\label{redshift}
\end{align}

Equation \eqref{redshift} demonstrates that the DMI causes a shift in the SW frequency that is linear in the DW velocity \cite{AFMDW-SW11,AFMDW-SW2}. This frequency shift significantly changes the interaction between SWs and DWs, especially in biaxial systems, where the angular frequency $\Omega$ is suppressed. A left-circularly (right-circularly) polarized SW incident on a DW has a positive (negative) frequency and induces a precession of the DW in the positive (negative) direction. In both cases, the effect of the angular frequency term, the first term in Eq. (\ref{redshift}), is the same: The precession of the DW reduces the frequency of the transmitted SW by the same amount for incoming SWs of the same amplitude but opposite helicity. By contrast, the velocity-dependent term, the second term in Eq. (\ref{redshift}), reinforces the redshift for left-circularly polarized SWs and reduces it for right-circularly polarized SWs.

Let us now derive the reactive magnonic force and torque. We employ the conservation of the energy-momentum tensor $T^{\alpha\beta}=g^{\alpha\gamma}T^{\beta}_{\gamma}$, where $g^{\alpha\gamma}=\mathrm{diag}\left(1,-1\right)$ is the inverse of the metric tensor in the Minkowski space $(x^0,x^1)=(t,x)$ and $T_{\alpha}^{\beta}=\partial_\alpha \bm{n}\cdot\partial \mathcal{L}/\partial(\partial_\beta \bm{n})-\delta_{\alpha}^{\beta} \mathcal{L}$. When Lagrangian density is invariant with respect to the space-time, the Noether's theorem implies continuity equations for the energy-momentum tensor $\partial_{\beta}T^{\alpha\beta}=0$, and current $\partial_{\alpha}j^\alpha=0$ \cite{spin-current}, respectively.

The translational and rotational symmetries in uniaxial AFM systems dictate the conservation of total linear-momentum $P$, and angular-momentum $J$.
Then the force and torque on AFM-CDWs are defined as $F=T^{11}(-\infty)-T^{11}(+\infty)=d{P}/dt$, and $\tau=j^{1}(-\infty)-j^{1}(+\infty)=d{J}/dt$, respectively.
The total reactive magnonic force $F=F_{\mathrm{reflection}}+F_{\mathrm{redshift}}+F_{\mathrm{DM}}$ consists of the reflection force, the redshift force, and the DMI force,
\begin{subequations}\label{reactiveforce}
\begin{align}
&F_{\mathrm{reflection}}=2|\Psi|^2|r|^2k_{-}^2 \label{force-ref},\\
&F_{\mathrm{redshift}}=|\Psi|^2(1-|r|^2)k_{-}(k_{-}-k_{+}), \label{force-red}\\
&F_{\mathrm{DM}}=2 \gamma D|\Psi|^2(1-|r|^2) k_{-}. \label{force-dmi}
\end{align}
\end{subequations}
Both the redshift and DMI forces are due to the transmitted SWs. The total reactive torque exerted by SWs is given by
\begin{align}
&\tau=2|\Psi|^2(1-|r|^2)k_{-}. \label{reactivetorque}
\end{align}
Equation (\ref{reactiveforce}) shows that when the SWs are hard ($k_\pm\gg 1$) and the DMI is large and comparable to the critical value $D\lesssim 1$, $F_{\mathrm{reflection}}\ll F_{\mathrm{redshift}}, F_{\mathrm{DMI}}$. If $k_+ \simeq k_-$, then the DMI force dominates ($F_{\mathrm{redshift}}\ll F_{\mathrm{DMI}}$). When $k_+ \ll k_-$, both the redshift and DMI forces are of the same order. For soft SWs ($k_\pm\ll 1$) and a strong redshift ($k_+\ll k_-$), the reflection force dominates.

%%%%%%%%%%%%%%%%%%%%%%%%%%%%%
%%%%%%%%%%%%%%%%%%%%%%%%%%%%

Thus far, we have not considered the ubiquitous dissipation. We include dissipative effects via a Rayleigh's dissipation function density $\mathcal{R}=\alpha (\partial_t\bm{n})^2/2$, where $\alpha$ is the Gilbert damping. We define a viscous force $F^v$ and a torque $\tau^v$ such that $d{P}/dt=F+F^v$ and $d{J}/dt=\tau+\tau^v$. In the steady state, $F=-F^v$ and $\tau=-\tau^v$, and we find that
\begin{subequations}\label{visc}
\begin{align}
&F=-\frac{2\alpha\gamma \left(D(\Omega-\gamma V D)-V (1-D^2-(\Omega-\gamma V D)^2)\right)}{\sqrt{1-D^2-(\Omega-\gamma V D)^2}}, \label{visc-torque}\\
&\tau=\frac{2\alpha\gamma(\Omega-\gamma VD)}{\sqrt{1-D^2-(\Omega-\gamma V D)^2}}. \label{vis-force}
\end{align}
\end{subequations}
Equations (\ref{ref-tran}) and (\ref{reactiveforce})-(\ref{visc}) form a closed set that we can solve numerically to find the steady state. We plot the results of the related simulations in Fig. \ref{k08}. We see that the directions of the angular frequencies of the DWs are opposite for different SW helicities.

We check our results based on Eqs. (\ref{ref-tran}) and (\ref{reactiveforce})-(\ref{visc}) against those of another, more direct numerical procedure. For the latter calculations, we follow Ref. \cite{AFMDW-SW1}. We solve the coupled nonlinear Landau-Lifshitz-Gilbert (LLG) equations for the staggered field $\bm{n}$ and the magnetization $\bm{m}$ \cite{AFMDW-SW1}. A circularly-polarized magnetic field of frequency $\omega_0$ and amplitude $h_0$ excites SWs in a region far from the DW center.
The amplitude of the external magnetic field is  in natural units.
\begin{figure}[t]
\includegraphics[width=8.35cm]{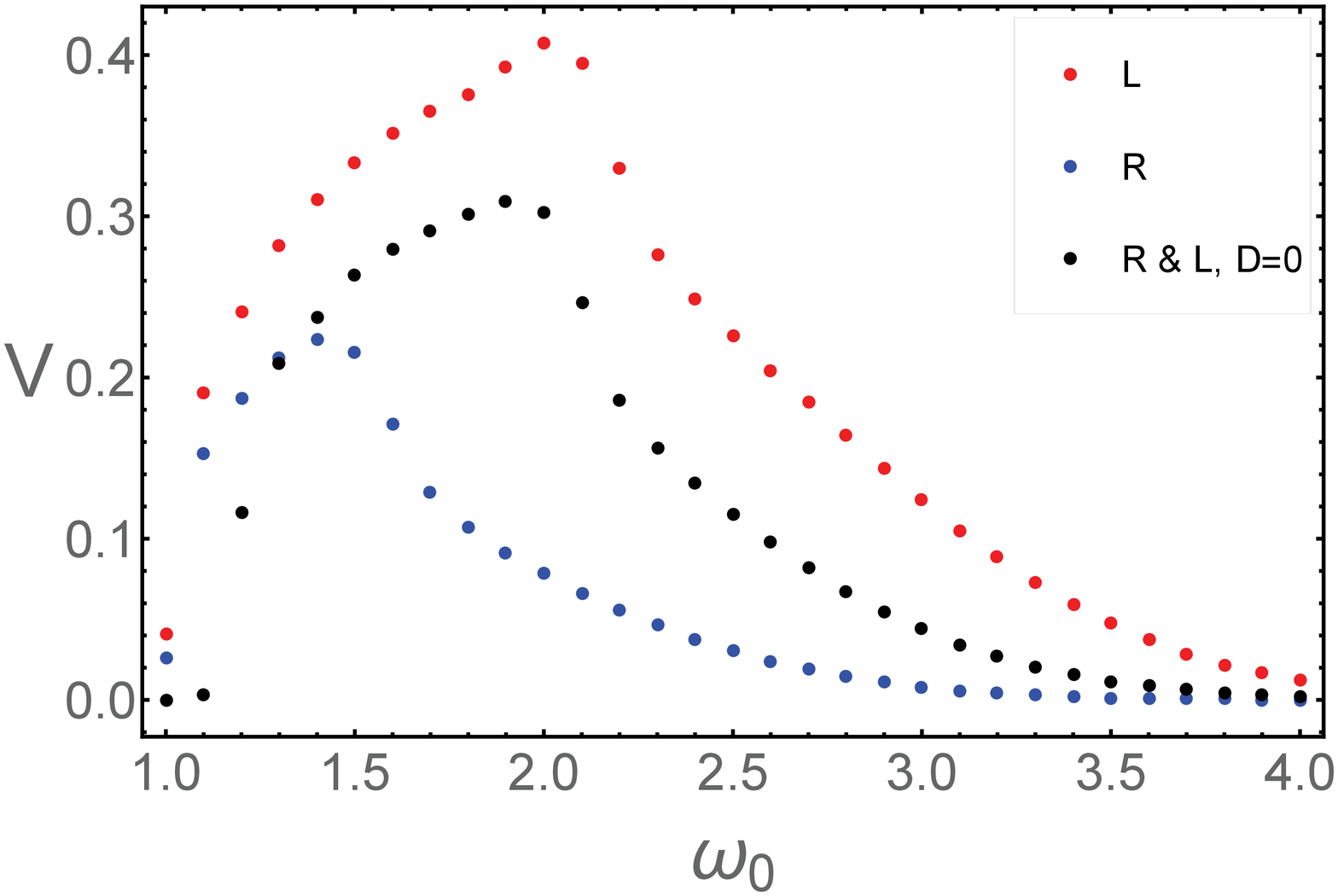}
\caption{The DW velocity $V$ as a function of the external magnetic field frequency, $\omega_0$, for both L- and R-polarized SWs in a uniaxial, $\kappa=0$, AFM system with $D=0.5$. The black dots represent the DW velocity in the absence of the DMI, $D=0$.}
\label{v-w}
\end{figure}
%%%%%%%%%%%%%%%%%%%%%%%%%%%%%

Figure \ref{v-w} shows the DW velocity as a function of $\omega_0$ for left- and right-circularly polarized SWs excited by a magnetic field of $h_0=0.001$. In the presence of a high frequency left-handed (right-handed) SW, the velocity of a chiral AFM-DW is higher (lower) than that of its non-chiral counterpart. The DW velocity exhibits non-monotonic behavior with respect to the frequency $\omega_0$. At lower frequencies, CDWs have higher velocities than their non-chiral counterparts for both helicities.

Note that these calculations are based on the conservation of linear momentum \cite{AFMDW-SW1,AFMDW-SW11,AFMDW-SW2} and they are valid for the small Gilbert damping regime \cite{Yaruslav}.

\textit{DW motion in a biaxial AFM system.---} We model a biaxial AFM material by introducing an additional contribution to the free-energy density expressed in Eq. (\ref{free-energy-n}), $U \rightarrow U + U_{\mathrm{ani}}$.  The transverse-axis anisotropy energy density is $U_{\mathrm{ani}}=\kappa(\bm{n} \cdot \hat{z})^2$, where $\kappa=K_z/2K_x > 0$ and $K_z$ is the anisotropy energy density in the $z$-direction. This fixes the CDW tilt angle such that $\Phi=0$ or $\pi$. For biaxial AFM systems, the spin-current $j^1$ is no longer conserved.
%%%%%%%%%%%%%%%%%%%%%%%%%%%%
\begin{figure}[t]
\includegraphics[width=8.5cm]{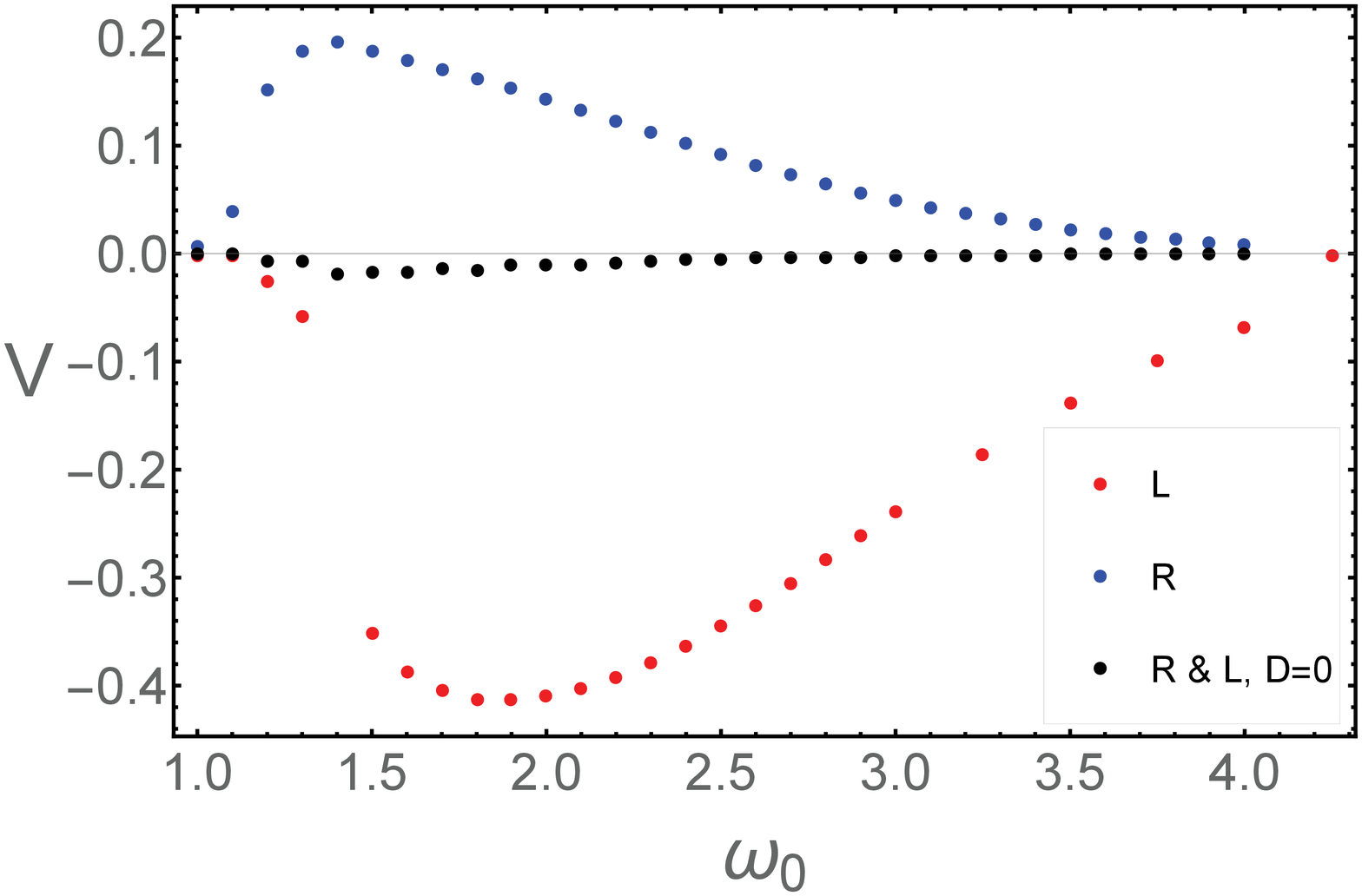}
\caption{The same as in Fig. \ref{v-w} but for a biaxial AFM system. The hard-axis anisotropy is $\kappa=0.25$.}
\label{v-w-hard}
\end{figure}
%%%%%%%%%%%%%%%%%%%%%%%%%%%%%

In a collinear biaxial AFM system, the SW polarization is elliptical with a dispersion relation:
\begin{align}
\omega_{\mathrm{b}}^2=1+\kappa+k^2\pm\sqrt{4 D^2 k^2+ \kappa^2}. \label{SW-col-bi}
\end{align}
In a biaxial AFM system, the spectrum remains symmetric, such that $\omega_{\mathrm{b}}(k)=\omega_{\mathrm{b}}(-k)$. Note that, even in the absence of the DMI, a transverse anisotropy lifts the helicity degeneracy of the polarized SWs; however, this degeneracy breaking is independent of momentum and leads to a finite gap between the two branches of the SW excitations.
The group velocity of the SWs in a collinear biaxial AFM system is  $v_g=\left(1\pm 2D^2/\sqrt{4D^2k^2+\kappa^2}\right)k/\omega_{\mathrm{b}}$. The equations of motion in biaxial AFM systems are complicated; here, we present only the results of numerical calculations.

In a biaxial AFM system, in the absence of the DMI, a DW approaches the SW source regardless of the SW helicity. There is no rotation of the DW. The potential in the SW Hamiltonian that is induced by the AFM-DW texture causes no reflections. Thus, the motion induced by elliptical SWs is similar to that induced by linearly polarized SWs in a uniaxial AFM system; the DW motion is slow compared with that induced by circularly polarized SWs and is toward the SW source because of the conservation of linear momentum \cite{AFMDW-SW1}.

In a biaxial AFM system, the DMI does not cause the CDWs to rotate, but Eq. (\ref{redshift}) shows that the outgoing SWs acquire a DMI-dependent frequency shift. The hard-axis anisotropy suppresses the angular frequency of the CDWs. The frequency shift simplifies to $\Delta\omega\simeq 2\gamma V D$. The DMI and redshift forces dominate the total magnonic force since the reflection force can be neglected. Figure \ref{v-w-hard} shows that DMI dramatically increases the DW velocity and leads to helicity-dependent DW motion. This is our second central result. The helicity of the SWs controls the direction of the induced motion of chiral AFM-DWs. The velocity of CDW is antisymmetric respect to the helicity of SWs. The DW velocity is approximately two times higher in the presence of one helicity (left-circularly polarized SWs, in our configuration) than it is in the presence of the opposite helicity.

\textit{Discussion and conclusion.---}
Using $\mathrm{KMnF_3}$ parameters \cite{parameters}, we find a DW velocity of $2~\mathrm{km/s}$ which is at least two orders of magnitude faster than the SW-driven CDW motion in biaxial FM systems \cite{FMDW-DMI} and one order of magnitude larger than the FM-DW driven by spin-orbit torque \cite{Miron}. On the other hand, although in SW-driven CDW motion only a change in the sign of the DMI merely reverses the direction of motion \cite{FMDW-DMI, Berakdar}, via a linear-momentum transfer mechanism \cite{Sinova}, in AFMs the SW helicity also controls the direction.

Recent developments in atomic-scale resolution microscopy enables us to trace the AFM-DW position via techniques, such as spin-polarized scanning-tunneling microscopy \cite{Bode}, photoelectron emission microscopy \cite{Weber}, and magnetic exchange force microscopy \cite{Roland}. The position of AFM-DW center also can be detected by using the anisotropic magneto resistance effect, indirectly \cite{Alireza1}.

The DMI results in a faster and more controlled motion of AFM-DWs. In a biaxial AFM system, the SW helicity determines both the direction and the magnitude of the DW velocity. These features enable the magnonic helicity-dependent DW motion. By contrast, in a uniaxial AFM system, a CDW always is recoiled from the SWs' source via a magnonic force with a helicity-dependent magnitude.

\section*{Acknowledgments}
The authors thank E. G. Tveten and Y. Tserkovnyak  for fruitful discussions. The research leading to these results received funding from the European Research Council via Advanced Grant No. 669442 ``Insulatronics" and was partially supported by the Research Council of Norway through its Centres of Excellence funding scheme, Project No. 262633, ``QuSpin".


\begin{thebibliography}{99}
\bibitem{energy-terms}
The parameters of the 1D micromagnetic model are defined in terms of the atomistic spin-lattice parameters as follow:
$\chi^{-1}=4 |J_0| S^2/a$, $A=2 a |J_0| S^2$, $K_x=2 K_{0} S^2/a$, $\bm{d}=2 \bm{D}_0 S^2/a$, and $\bm{D}=\bm{D}_0 S^2$, where $J_0$ and $K_0$ are the Heisenberg exchange energy and anisotropy energy, respectively, and $a$ is the lattice constant.
\bibitem{topological-term}
In our non linear $\sigma$ model, we disregarded the topological term [$\chi L \dot{\bm{n}} \cdot (\bm{\nabla}\bm{n}\times \bm{n})$] and another term arising from the homogeneous DMI [$\chi \dot{\bm{n}} \cdot \bm{d}$], because these terms are total derivatives and do not affect the equations of motion.
\bibitem{SW-expression}
The staggered field can be expressed in terms of $\psi$:
$\bm{n}=\mathrm{Re}[\psi] \hat{\mathrm{e}}_1+\mathrm{Im}[\psi]  \hat{\mathrm{e}}_2+\left[1-(\mathrm{Re}[\psi])^2/2-(\mathrm{Im}[\psi])^2/2 \right]\hat{\mathrm{e}}_3$.
\bibitem{spin-current}
The conserved spin-currents are expressed as $j^\alpha=n_y \partial\mathcal{L}/\partial(\partial_\alpha n_z)-n_z \partial\mathcal{L}/\partial(\partial_\alpha n_y)$.
\bibitem{Yaruslav}
In the presence of larger damping constants another mechanism becomes dominant in uniaxial AFMs.
In this case SWs decay quickly thus there are much more SWs on one side of AFM-DW than on the other side, consequently this leads to a large gradient of the effective stiffness field amplitude. This effective stiffness field gradient leads to a backward motion of AFM-DWs in the presence of circularly polarized SWs in systems with uniaxial anisotropy. This is similar to the contribution of so-called entropic torque (or force) on the DW motion in the presence of a temperature gradient. The entropic torque arises from the temperature dependence of the effective exchange field as discussed in Ref \cite{Yaruslav1}.
\bibitem{parameters}
We use the material parameters for $\mathrm{KMnF_3}$ in 1D: $S=5/2$, $a=0.418$nm, $M_s=5\times10^{-14}$Am, $K_x=2.03\times10^{-14}$J/m, $D=1 \times 10^{-22}$J,  and $A=1.15\times 10^{-30}$ Jm, adapted from Ref. \cite{uniaxialAFM}. Also we consider $\alpha=10^{-4}$. Other candidate materials for the observation of SW-induced AFM-DW motion include metallic AFM materials such as the uniaxial $\mathrm{KMnF_3}$ \cite{uniaxialAFM} and insulating AFM materials such as the uniaxial $\mathrm{Cu_2V_2O_7}$ \cite{AFMSW4-DMI} as well as the biaxial $\mathrm{Sr_2IrO_4}$ and $\mathrm{Ba_2IrO_4}$ \cite{biaxialAFM}.
\end{thebibliography}
\end{document}